\documentclass[a4paper,11pt]{article}
\usepackage[utf8]{inputenc}
\usepackage[catalan, english]{babel}
\usepackage{geometry}
\usepackage{graphicx}
\graphicspath{{./figures/}}
\usepackage{natbib}
\usepackage{appendix}
\usepackage{tabularx}
\usepackage[catalan,english]{varioref}
\vrefwarning
\usepackage{color}
\usepackage{booktabs}
\usepackage{threeparttable}
\usepackage{longtable}
\usepackage{multicol}
\usepackage{multirow}
\usepackage{enumerate}
\usepackage{amsmath}
\usepackage{epstopdf}
\usepackage{subcaption}
\usepackage{amssymb}
\usepackage{setspace} 
\usepackage{hyperref}
\usepackage[english]{cleveref}
\usepackage[toc,acronym]{glossaries} 
\usepackage{float}
\definecolor{halfgray}{gray}{0.55}
\definecolor{naviblue}{RGB}{0,0,102}
\definecolor{webbrown}{rgb}{.6,0,0}
\definecolor{RoyalBlue}{cmyk}{1, 0.50, 0, 0}
\definecolor{webgreen}{rgb}{0,.5,0}
\definecolor{Maroon}{cmyk}{0, 0.87, 0.68, 0.32}
\definecolor{Black}{cmyk}{0, 0, 0, 0}
\definecolor{myorange}{RGB}{239, 186,67}

\hypersetup{%
    colorlinks=true,linktoc=all, pdfstartpage=1, pdfstartview=FitV,%
    breaklinks=true, pdfpagemode=UseNone, pageanchor=true, pdfpagemode=UseOutlines,%
    plainpages=false, bookmarksnumbered, bookmarksopen=true, bookmarksopenlevel=2,%
    hypertexnames=true, pdfhighlight=/O,
    urlcolor=webbrown, linkcolor=RoyalBlue, citecolor=webgreen,
}

\newcommand{\lya}{Ly$\alpha$}

\newcommand{\cm}{\,\mathrm{cm}}

\newcommand{\kms}{\, {\rm km}\, {\rm s}^{-1} }

\newcommand{\hmpc}{\, h^{-1} {\rm Mpc}}

\newcommand{\msun}{\, {\rm M_\odot} }

\crefname{equation}{equation}{equations}
\crefname{figure}{figure}{figures}
\crefname{appsec}{appendix}{appendices}
\creflabelformat{equation}{#2#1#3}

\glsaddkey
 {descripcio}
 {\relax}
 {\glsentrydescripcio}
 {\Glsentrydescripcio}
 {\glsdescripcio}
 {\Glsdescripcio}
 {\GLSdescripcio}
\glsaddkey
 {descripcioplural}
 {\relax}
 {\glsentrydescripcioplural}
 {\Glsentrydescripcioplural}
 {\glsdescripcioplural}
 {\Glsdescripcioplural}
 {\GLSdescripcioplural}

\makeglossaries
\setacronymstyle{long-short}
\glssetnoexpandfield{first}
\glssetnoexpandfield{firstpl}

\newglossaryentry{2dF}{
	type={acronym},
	name={2dF Survey},
	description={Two-degree-Field Galaxy Redshift survey},	
	first={\glsentrytext{2dF}},
}

\newglossaryentry{AGN}{
	type={acronym},
	name={AGN},
	description={Active Galactic Nuclei},
	descripcio={galàxia de nucli actiu},
	first={\glsentrydesc{AGN} (\glsentrytext{AGN})},
}

\newglossaryentry{BAL}{
	type={acronym},
	name={BAL},
	description={Broad Absorption Line},	
	first={\glsentrydesc{BAL} (\glsentrytext{BAL})},
	plural={BALs},
	descriptionplural={Broad Absorption Line},
	firstplural={\glsentrydescplural{BAL} (\glsentryplural{BAL})},
}

\newglossaryentry{BAO}{
	type={acronym},
	name={BAO},
	description={Baryon Acoustic Oscillations},
	descripcio={oscil·lacions acústiques dels barions},
	first={\glsentrydesc{BAO} (\glsentrytext{BAO})},
}

\newglossaryentry{BBN}{
	type={acronym},
	name={BBN},
	description={Big Bang Nucleosynthesis},
	descripcio={nucleosíntesi primordial},
	first={\glsentrydesc{BBN} (\glsentrytext{BBN})},
}

\newglossaryentry{BI}{
	type={acronym},
	name={BI},
	description={balnicity index},
	first={\glsentrydesc{BI} (\glsentrytext{BI})},
}

\newglossaryentry{BOSS}{
	type={acronym},
	name={BOSS},
	description={Baryon Oscillations Spectroscopic Survey},
	first={\glsentrydesc{BOSS} (\glsentrytext{BOSS})},
}

\newglossaryentry{CDM}{
	type={acronym},
	name={CDM},
	description={Cold Dark Matter},
	descripcio={matèria fosca freda},
	first={\glsentrydesc{CDM} (\glsentrytext{CDM})},
}

\newglossaryentry{CFHTLenS}{
	type={acronym},
	name={CFHTLenS},
	description={a lensing catalogue},
	first={\glsentrytext{CFHTLenS}},
}

\newglossaryentry{CMASS}{
	type={acronym},
	name={CMASS},
	description={a catalogue of ``constant mass" luminous infrared galaxies from \gls{BOSS}},
	first={\glsentrytext{CMASS}},
}

\newglossaryentry{CMB}{
	type={acronym},
	name={CMB},
	description={Cosmic Microwave Background},
	descripcio={fons còsmic de microones},
	first={\glsentrydesc{CMB} (\glsentrytext{CMB})},
}

\newglossaryentry{CNR}{
	type={acronym},
	name={CNR},
	description={continuum-to-noise ratio},
	first={\glsentrydesc{CNR} (\glsentrytext{CNR})},
}

\newglossaryentry{COBE}{
	type={acronym},
	name={COBE},
	description={COsmic Background Explorer},
	first={\glsentrytext{COBE}},
}

\newglossaryentry{DE}{
	type={acronym},
	name={DE},
	description={dark energy},
	first={\glsentrydesc{DE} (\glsentrytext{DE})},
}

\newglossaryentry{DLA}{
	type={acronym},
	name={DLA},
	description={Damped \lya{} Absorber},
	descripcio={sistema \lya{} esmorteïta},
	descripcioplural={sistemes \lya{} esmorteïts},
	descriptionplural={Damped \lya{} Absorbers},
	first={\glsentrydesc{DLA} (\glsentrytext{DLA})},	
	plural={DLAs},
	firstplural={\glsentrydescplural{DLA} (\glsentryplural{DLA})},
}

\newglossaryentry{DM}{
	type={acronym},
	name={DM},
	description={dark matter},
	first={\glsentrytext{DM}},
}

\newglossaryentry{DR}{
	type={acronym},
	name={DR},
	description={Data Release},
	first={\glsentrydesc{DR} (\glsentrytext{DR})},
}

\newglossaryentry{DRXQ}{
	type={acronym},
	name={DRXQ},
	description={\ifglsused{DR}{DRX quasar catalogue}{Data Release 9 quasar catalogue\glsunset{DR}}},
}

\newglossaryentry{DR7}{
	type={acronym},
	name={DR7},
	description={seventh Data Release},
	first={\ifglsused{DR}{\glsentrytext{DR7}}{\glsentrydesc{DR7} (\glsentrytext{DR7}) \glsunset{DR}}},
}

\newglossaryentry{DR9}{
	type={acronym},
	name={DR9},
	description={ninth Data Release},
	first={\ifglsused{DR}{\glsentrytext{DR9}}{\glsentrydesc{DR9} (\glsentrytext{DR9}) \glsunset{DR}}},
}

\newglossaryentry{DR9Q}{
	type={acronym},
	name={DR9Q},
	description={DR9 quasar catalogue},
	first={\ifglsused{DRXQ}{\glsentrytext{DR9Q}}{\ifglsused{DR}{\glsentrydesc{DR9Q} (\glsentrytext{DR9Q})}{\glsentrydesc{DR9} quasar catalogue (\glsentrytext{DR9Q})}\glsunset{DRXQ}\glsunset{DR}}},
}
	
\newglossaryentry{DR10}{
	type={acronym},
	name={DR10},
	description={tenth Data Release},
	first={\ifglsused{DR}{\glsentrytext{DR10}}{\glsentrydesc{DR10} (\glsentrytext{DR10}) \glsunset{DR}}},
}

\newglossaryentry{DR11}{
	type={acronym},
	name={DR11},
	description={eleventh Data Release},
	first={\ifglsused{DR}{\glsentrytext{DR11}}{\glsentrydesc{DR11} (\glsentrytext{DR11}) \glsunset{DR}}},
}

\newglossaryentry{DR12}{
	type={acronym},
	name={DR12},
	description={twelfth Data Release},
	first={\ifglsused{DR}{\glsentrytext{DR12}}{\glsentrydesc{DR12} (\glsentrytext{DR12}) \glsunset{DR}}},
}

\newglossaryentry{DR12Q}{
	type={acronym},
	name={DR12Q},
	description={DR12 quasar catalogue},
	first={\ifglsused{DRXQ}{\glsentrytext{DR12Q}}{\ifglsused{DR}{\glsentrydesc{DR12Q} (\glsentrytext{DR12Q})}{\glsentrydesc{DR12} quasar catalogue (\glsentrytext{DR12Q})}\glsunset{DRXQ}\glsunset{DR}}},
}

\newglossaryentry{FLRW}{
	type={acronym},
	name={FLRW},
	description={Friedmann–Lema\^itre–Robertson–Walker},
	first={\glsentrydesc{FLRW} (\glsentrytext{FLRW})},
}

\newglossaryentry{FRW}{
	type={acronym},
	name={FRW},
	description={Friedmann–Robertson–Walker},
	first={\glsentrydesc{FRW} (\glsentrytext{FRW})},
}

\newglossaryentry{GR}{
	type={acronym},
	name={GR},
	description={general relativity},
	descripcio={relativitat general},
	first={\glsentrydesc{GR} (\glsentrytext{GR})},
}
	
\newglossaryentry{HCD}{
	type={acronym},
	name={HCD},
	description={high-column-density system},
	descriptionplural={high-column-density systems},
	first={\glsentrydesc{HCD} (\glsentrytext{HCD})},	
	plural={HCDs},
	firstplural={\glsentrydescplural{HCD} (\glsentryplural{HCD})},
}
	
\newglossaryentry{IGM}{
	type={acronym},
	name={IGM},
	description={intergalactic medium},
	descripcio={medi intergalàctic},
	first={\glsentrydesc{IGM} (\glsentrytext{IGM})},
}

\newglossaryentry{LCDM}{
	type={acronym},
	name={$\Lambda$CDM},
	description={Lambda Cold Dark Matter},
	first={\ifglsused{CDM}{\glsentrytext{LCDM}}{\glsentrydesc{LCDM} (\glsentrytext{LCDM}) \glsunset{CDM}}
},
}

\newglossaryentry{LSS}{
	type={acronym},
	name={LSS},
	description={large-scale structure},
	first={\glsentrydesc{LSS} (\glsentrytext{LSS})},
}

\newglossaryentry{MJD}{
	type={acronym},
	name={MJD},
	description={Modified Julian Date},
	first={\glsentrydesc{MJD} (\glsentrytext{MJD})},
}

\newglossaryentry{MTC}{
	type={acronym},
	name={MTC},
	description={Mean Transmission Correction },
	first={\glsentrydesc{MTC} (\glsentrytext{MTC})},
}

\newglossaryentry{PSF}{
	type={acronym},
	name={PSF},
	description={Point Spread Function},
	first={\glsentrydesc{PSF} (\glsentrytext{PSF})},
}

\newglossaryentry{RMS}{
	type={acronym},
	name={RMS},
	description={root mean square},
	first={\glsentrydesc{RMS} (\glsentrytext{RMS})},
}

\newglossaryentry{SDSS}{
	type={acronym},
	name={SDSS},
	description={Sloan Digital Sky Survey},
	first={\glsentrydesc{SDSS} (\glsentrytext{SDSS})},
}

\newglossaryentry{SDSS-II}{
	type={acronym},
	name={SDSS-II},
	description={Sloan Digital Sky Survey II},
	first={\ifglsused{SDSS}{\glsentrytext{SDSS-II}}{\glsentrydesc{SDSS-II} (\glsentrytext{SDSS-II})  \glsunset{SDSS}}},
}

\newglossaryentry{SDSS-III}{
	type={acronym},
	name={SDSS-III},
	description={Sloan Digital Sky Survey III},
	first={\ifglsused{SDSS}{\glsentrytext{SDSS-III}}{\glsentrydesc{SDSS-III} (\glsentrytext{SDSS-III})  \glsunset{SDSS}}},
}

\newglossaryentry{SN}{
	type={acronym},
	name={SN},
	description={supernova},
	descriptionplural={supernovae},
	descripcio={supernova},
	descripcioplural={supernoves},
	first={\glsentrydesc{SN} (\glsentrytext{SN})},
	plural={SNe},
	firstplural={\glsentrydescplural{SN} (\glsentryplural{SN})},
}

\newglossaryentry{WMAP}{
	type={acronym},
	name={WMAP},
	description={Wilkinson Microwave Anisotropy Probe},
	first={\glsentrytext{WMAP}},
}

\begin{document}

\title{The Cosmological Bias Factor of Damped Lyman Alpha systems:
 Dependence on Metal Line Strength}
\author{
Ignasi P\'erez-R\`afols$^{1}$, Jordi Miralda-Escud\'e$^{2,3}$, Andreu Arinyo-i-Prats$^{3}$, \\
Andreu Font-Ribera$^{4}$, Lluís Mas-Ribas$^{5}$
}
\date{\flushleft{
\footnotesize
$^{1}$Aix Marseille Univ, CNRS, CNES, LAM, Marseille, France\\
$^{2}$Instituci\'o Catalana de Recerca i Estudis Avan\c{c}ats, Barcelona, Catalonia\\
$^{3}$Institut de Ci\`encies del Cosmos, Universitat de Barcelona/IEEC, Barcelona, E-08028, Catalonia\\
$^{4}$Department of Physics and Astronomy, University College London, Gower Street, London, United Kingdom\\
$^{5}$Institute of Theoretical Astrophysics, University of Oslo, P.O. Box 1029 Blindern, NO-0315 Oslo, Norway\\}}

\maketitle

\begin{abstract}
  \glsunset{DLA}

  We measure the cosmological bias factor of \glspl{DLA} from their
cross-correlation with the \lya{} forest absorption, as a function of the
\gls{DLA} metal strength, defined from an average of equivalent widths
of the strongest detectable low-ionization metal lines. A clear increase
of the bias factor with metal strength is detected, as expected from a
relation of metallicity and velocity dispersion with host halo mass. The
relation is stronger after the metal strength is corrected for the HI
column density, to make it more related to metallicity instead of metal
column density. After correcting for the effects of measurement errors
of the metal strength parameter, we find that the bias factor of \glspl{DLA}
with the weakest metal lines is close to unity, consistent with an
origin in dwarf galaxies with host halo masses $\sim 10^{10}\msun$,
whereas the most metal rich \glspl{DLA} have a bias factor as large as
$b_{\rm DLA}\sim 3$, indicative of massive galaxies or galaxy groups in
host halos with masses $\sim 10^{12}\msun$. Our result confirms
the physical origin of the relation of bias factors measured from
cross-correlation studies to the host halos of the absorbers.

\glsreset{DLA}

\end{abstract}

\tableofcontents

%
\section{Introduction}  \label{sec DLA: Introduction}

 The \lya{} forest is a fluctuating absorption seen bluewards of the
\lya{} rest-frame wavelength of the source caused by intergalactic
hydrogen. Generally, regions of higher density give rise to absorption
features of higher hydrogen column density. When the column density is
as high as $N_{\rm HI}\ge 2\times10^{20}\cm^{-2}$ the hydrogen becomes
self-shielded against the external cosmic ionizing background, and
the gas is mostly in atomic form. These systems are called\glspl{DLA} \citep{Wolfe+1986}, and their damped profiles are measurable even in low
resolution and low signal-to-noise spectra, providing a robust method to
measure their column densities.
The contribution of these systems to the cosmic density of atomic gas
is $\Omega_{\rm DLA}\simeq10^{-3}$ at redshifts $2<z<3.5$
\citep{Peroux+2003,Prochaska+2005,Zafar+2013,Crighton+2015,
Padmanabhan+2016,Prochaska+2009,Noterdaeme+2009,Noterdaeme+2012b},
corresponding to $\sim 2\%$ of all the baryons in the Universe.
Absorbers of this high column density naturally arise in halos where gas
cools and forms dense clouds, and are therefore crucial to understand
the galaxy formation process from gas that is accreted into halos and expelled in galactic winds.

  Even though the contribution of the \glspl{DLA} to the matter density
is well understood, the sizes of galaxies and the masses of halos
hosting the absorbing gas cannot be inferred from the properties of the
absorption lines, and it was widely believed until recently that the
majority of \glspl{DLA} were hosted in dwarf galaxies and low-mass
halos. One of the avenues to determine the characteristic masses of
\gls{DLA} host halos is to use the cross-correlation of \glspl{DLA} with
other tracers of large-scale structure to measure their cosmological bias
factor. The standard \gls{CDM} model of structure formation
predicts the bias as a function of halo mass: the larger the halo mass,
the larger the bias factor. This has been exploited by means of the
cross-correlation with the \lya{} forest transmission fluctuations
by \cite{Font-Ribera+2012,Perez-Rafols+2018}. The most recent value of
the bias factor obtained from the analysis of the final \gls{DR12} of the \gls{BOSS} survey from \gls{SDSS-III} is $b_{\rm DLA}=1.97\pm 0.08$,
implying substantially higher halo masses than previously believed.
Models where \glspl{DLA} are present in a broad range of halo masses,
including dwarf galaxies but also massive galaxies and galaxy groups,
are consistent with this value of the bias factor if the average halo
cross section to produce a \gls{DLA} increases with the halo mass $M_h$
at least as steeply as $\Sigma_{\rm DLA}\propto M_h$. The results also
show there is no dependence of the \gls{DLA} bias on hydrogen column
density, and no evolution with redshift to within measurement errors.

  There are other observable properties of \glspl{DLA}, however, that we
expect the bias factor to depend on. The velocity dispersion of matter
in dynamical equilibrium in the halo should increase with the host halo
mass as $\sigma\propto M_h^{1/3}$ for a fixed halo collapse epoch.
The mass-metallicity relation observed between the stellar mass and the
metal abundance of stars, already in place at redshifts $z\sim 3$
\citep{Maiolino+2008}, suggests that a corresponding relation probably
holds between the metallicity in the gas phase and the host halo mass.
The strength of metal absorption lines should increase with metallicity,
and also with the velocity dispersion of the absorbing gas for saturated
lines. Therefore, the bias factor should increase with the strength of
metal lines.

  Studies of these metal lines have shown that DLA metallicities are
typically low and distributed over a broad range, and that the mean
metallicity decreases with redshift
\citep{Kulkarni+2002,Vladillo2002,
Prochaska+2003a,Kulkarni+2005,Rafelski+2012,Rafelski+2014,Jorgenson+2013,
Neeleman+2013}.
The gas velocity dispersion
can be measured when several absorption components are seen
\citep{Prochaska+1997,Prochaska+2008}, although their relation with a
halo velocity dispersion is uncertain and may be affected by disk
dynamics. A relation of metallicity and velocity dispersion has been
found from these observations 
\citep{Neeleman+2013}. These detailed
studies, however, can only be done with high resolution and
signal-to-noise spectra, which are not possible to obtain for the large
samples of \glspl{DLA} that are required to detect the large-scale
cross-correlation with the \lya{} forest.

  The only \gls{DLA} sample that is large enough at present to allow for an
accurate determination of the \gls{DLA}-\lya{} forest cross-correlation is the
one obtained from the \gls{BOSS} survey
\citep{Noterdaeme+2009,Noterdaeme+2012b}. In the \gls{BOSS} spectra, metal lines
are practically unresolved (the characteristic width of the broadest \gls{DLA}
systems is comparable to the \gls{BOSS} spectrograph resolution), and the
signal-to-noise is most often below $\sim 3$. These spectra allow only
a detection of metal lines and rough measurements of their equivalent
widths, although many average \gls{DLA} properties can be derived by stacking
the absorption spectra of many \glspl{DLA} \citep{Mas-Ribas+2017}.

  In this paper, we will use an estimate of the metal strength of
individual \glspl{DLA} obtained by averaging the equivalent widths of several
metal lines, which we have previously defined and studied in
\cite{Arinyo-i-Prats+2018}, to classify the \glspl{DLA} into several groups
of different metal strength. We will then measure the bias factor for
these groups to look for a dependence on the metal strength. It is not
possible in general to tell the dependence of the metal strength (or
equivalent widths of the strongest observable lines) on the metal
abundance and the velocity dispersion of the absorbing gas in each
individual \gls{DLA} without obtaining spectra of much higher quality. In
a future paper, we plan to investigate the average relation of this
metal strength parameter to metallicity and velocity dispersion
from the study of stacked \gls{DLA} absorption spectra, following the
technique of \cite{Mas-Ribas+2017}. For now, we simply note that
metallicity and velocity dispersion should be correlated with each
other and should both increase with host halo mass.

  We start by describing the datasets used to derive the \gls{DLA} bias
in \cref{sec DLA: Sample data}. The estimator for the
cross-correlation as well as the model used to derive the \gls{DLA} bias are described in \cref{sec DLA: Cross-correlation}. The results are presented and discussed in \cref{sec DLA: Results}, and we summarize our conclusions
in \cref{sec DLA: Conclusions}. Throughout this paper we use a flat \gls{LCDM} cosmology, with
$\Omega_{m}=0.3156$, $\Omega_{b}=0.0492$, $h=0.6727$, $n_{s}=0.9645$,
and $\sigma_{8}=0.831$, as reported by \cite{Planck2015}.

%
\section{Data Sample}  \label{sec DLA: Sample data}

 As in \cite{Perez-Rafols+2018}, we use the \gls{DLA} catalogue obtained with
the technique of \cite{Noterdaeme+2009} from the final \gls{DR12}
of \gls{BOSS} \citep{Dawson+2013}, from \gls{SDSS-III}
\citep{Gunn+1998,York+2000,Gunn+2006,Eisenstein+2011,Bolton+2012,Smee+2013}.
These \glspl{DLA} were searched in the final \gls{DR12} quasar catalogue
\cite{Paris+2017} that used the quasar target selection of \gls{BOSS},
as summarized in \cite{Ross+2012}.

  The \lya{} forest absorption, used as tracer of the underlying
density field around \glspl{DLA}, is measured from the same set of 
157,922 quasar spectra as in \cite{Perez-Rafols+2018}, which have
$\sim$ 27 million pixels in the \lya\ forest. Because the
wavelength resolution of the BOSS spectra is better than required to
measure the cross-correlation function at the scales we are interested
in, we actually use {\it analysis pixels} in all of our calculations,
described in \cite{Busca+2013}, which are the average of every three
pixels of the actual co-added spectra and have a width
$\Delta_v= (\Delta \lambda/\lambda) c \simeq 207\kms$.

  Our \gls{DLA} sample is defined starting from sample $C1$ of
\cite{Perez-Rafols+2018}, which includes 23,342 \gls{DLA} candidates
with column density $N_{\rm HI}\ge10^{20}\cm^{-2}$. Note that we use a
column density threshold lower than the standard \gls{DLA} threshold 
of $N_{\rm HI}\ge 2\times10^{20}\cm^{-2}$. The reason is that the
larger number of \glspl{DLA} we obtain implies a smaller statistical
error, and that no evidence of a change of the bias factor with column
density was found in \cite{Perez-Rafols+2018}. This threshold is in any
case a conventional definition since the transition to a mostly atomic
and self-shielded medium with increasing $N_{\rm HI}$ is a gradual one.
This $C1$ sample is drawn from an early version of the \gls{DR12}
extension of the \gls{DLA} catalogue from \cite{Noterdaeme+2012b}, after
applying the first three cuts described in \cite{Perez-Rafols+2018},
which are:
  \begin{enumerate}  
	\item The \gls{DLA} redshift is in the range $2.0\le z_{\rm DLA}<3.5$.
	\item We require an average \gls{CNR} in the \lya{} forest region
 $CNR \ge 3$.
	\item We exclude \glspl{DLA} found in quasar spectra with
positive Balnicity Index, as listed in the DR12Q catalogue of
\cite{Paris+2014}.
  \end{enumerate}  
The additional cuts 4 to 6 described in \cite{Perez-Rafols+2018} are not
applied here, because they were found not to significantly affect the
measured bias factor.

 The goal of this paper is to measure the dependence of the \gls{DLA}
bias factor on the metal strength parameter $S$, as defined in
\cite{Arinyo-i-Prats+2018}. Briefly, the $S$ parameter is an average of
the equivalent widths of metal lines associated with a \gls{DLA}, optimally
weighted to obtain the best possible signal-to-noise, and normalized so
that $S=1$ represents an average of the \gls{DLA} metal strength. The metal
strength $S$ is therefore a quantity that depends both on the column
density of the measured metal species and the velocity dispersion of the
\glspl{DLA}, because of line saturation effects on the measured equivalent
widths. The precise relation of S to the metal column
density and velocity dispersion is uncertain, and can be constrained
from the stacked spectra of \glspl{DLA} in different intervals of $S$, which we
plan to study in another paper. We note that the $S$ parameter depends
on the \gls{DLA} sample that is used, because it is normalized to have a
mean value of 1 in the sample, and is therefore a measure of the metal
strength compared to the mean of other \glspl{DLA}. However, our \gls{DLA} sample
is a representative set of absorbers intercepted on random lines of sight,
so the quantity $S$ is an actual estimate of the metal line strength
compared to the average \gls{DLA} in the universe.

The true value of $S$ should be positive for every \gls{DLA}, but
the frequently large spectral noise may occasionally render it negative.
Cases of \glspl{DLA}
with negative $S$ are also included to avoid biasing our sample. 
 There are a small number of \glspl{DLA} for which it is not possible
to measure the equivalent width of any line for various reasons
\citep[see][for more details]{Arinyo-i-Prats+2018}. Removing these
objects reduces our sample to 23,312 \glspl{DLA}.

  To measure the dependence of the bias with the metal strength, we
separate our \gls{DLA} sample into three bins in $S$, defined as
$S<0.55$, $0.55 \le S < 1.45$, and $1.45\le S$,
chosen to have similar numbers of \glspl{DLA} in each bin. We label the
sub-samples of \glspl{DLA} in each bin as $S1$, $S2$, and $S3$, as shown in
\cref{ta DLA: sample properties}. We estimate for each \gls{DLA} the error
of the value of $S$, $\epsilon_S$, measured from the equivalent widths
of several metal lines, as described in detail in
\cite{Arinyo-i-Prats+2018}. To avoid having too many objects that are
placed in the wrong bin because of the measurement error $\epsilon_S$,
we require this error to be smaller than a threshold value, which we
set for our standard case to be $\epsilon_S < 0.5$, although we shall
examine the dependence of our results on this value. In addition we
remove any \glspl{DLA} with a value $S < -2$, because we have found that these
systems suffer from systematic errors with a highly non-Gaussian
distribution tail. After removing all the \glspl{DLA} with an error
$\epsilon_S > 0.5$ or $S< -2$, our total sample is reduced to the $SA$
sample, with 18,221 \glspl{DLA}, of which roughly 6,000 are in each of the
three sub-samples $S1$, $S2$, and $S3$ (see \cref{ta DLA: sample properties}).
Note that the values of $S$ are defined to have a
mean value of 1 for our whole C1 sample, and that removing some objects
affects this mean. We do not renormalise the values of $S$, as seen in
table \ref{ta DLA: sample properties} (e.g., sample $SA$ has a mean $S$
slightly larger than 1).

\begin{figure*}
	\centering
	\includegraphics[width=0.5\textwidth]{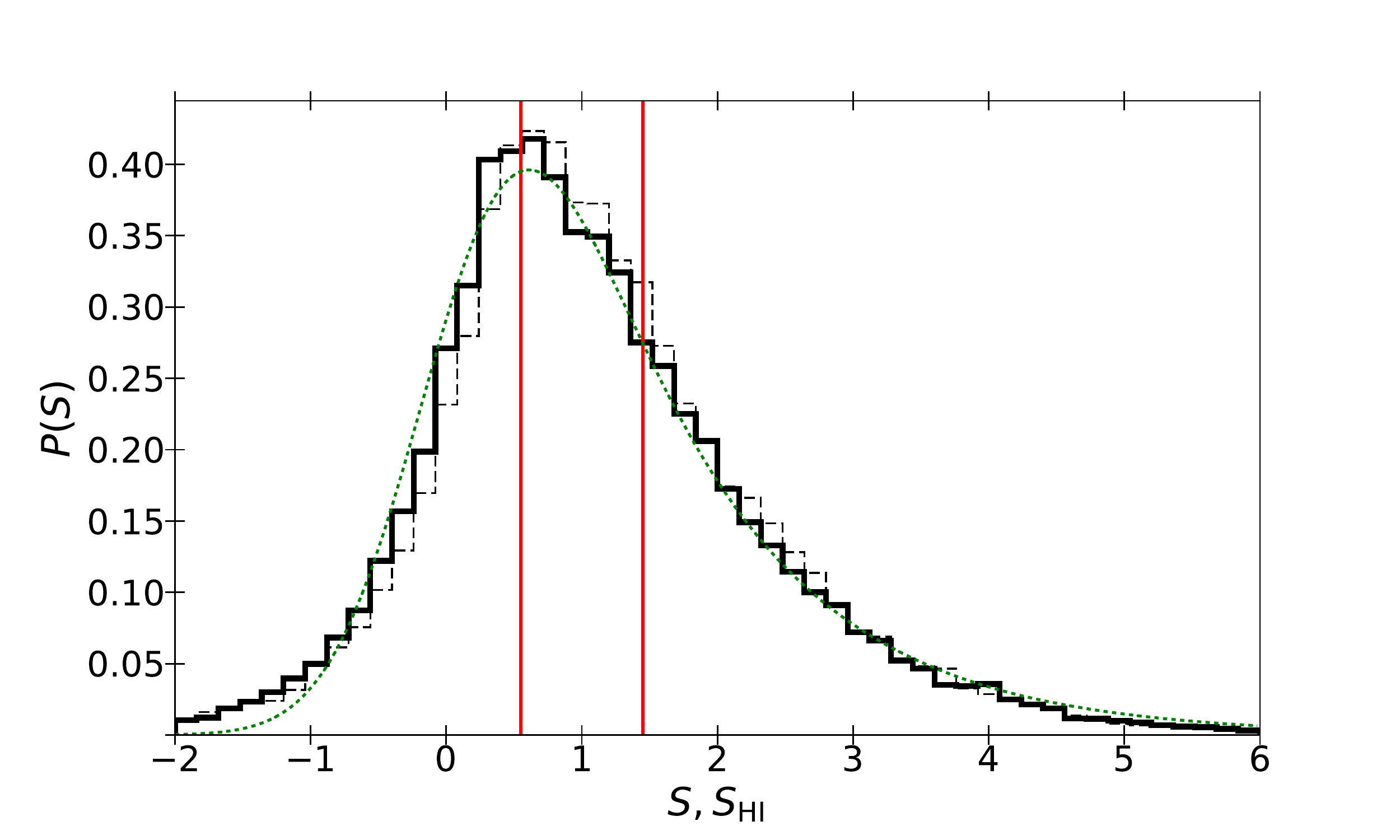}
	\hspace{-0.05\textwidth}
	\includegraphics[width=0.5\textwidth]{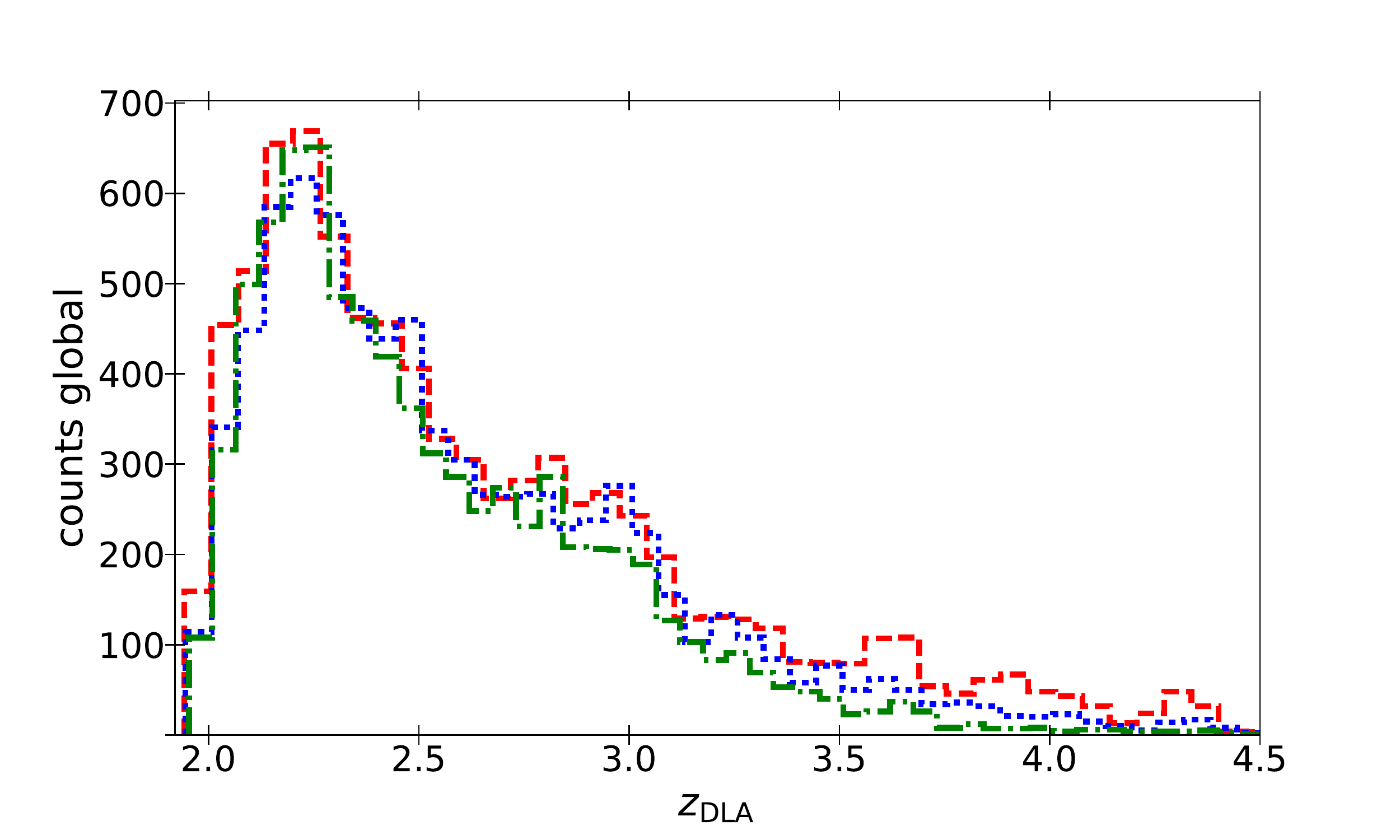}
	\caption{Left: Distribution of the metal strength $S$ for the
\glspl{DLA} in sample $SA$ (thick black line), and of the HI-corrected
metal strength $S_{\rm HI}$ (thin dashed line). Vertical red lines
indicate the bins used to construct the sub-samples $S1$ to $S3$. Green
dotted line shows an exponential profile with $\lambda_{S} = 1.2$
convolved with a Gaussian with error $\epsilon_{S} = 0.6$ (see
\cref{sec: error smoothing correction}), which adequately describes the
measured distribution. Right: Distribution of $z_{\rm DLA}$ for the
\glspl{DLA} in samples $S1$ (red dashed line), $S2$ (blue dotted line),
and $S3$ (green dashed-dotted line).}
	\label{fig DLA: dla hist}
\end{figure*}

  The left panel of \cref{fig DLA: dla hist} shows the distribution of
the $SA$ \gls{DLA} sample in metal strength. The bins separating the
sub-samples $S1$ to $S3$ are indicated as red solid lines. The right panel
shows the redshift distribution of sub-samples $S1$, $S2$, and $S3$. The redshift distributions are similar, although
there is an excess of low $S$ systems at high redshift. This excess can
be due to the known increase of metallicity with cosmic time, but also by
an increased contamination of false \glspl{DLA} which have weak or absent metal lines). Any at high redshift. In any case, this
small difference in the redshift distribution should not significantly
affect differences in the bias factor of the 3 sub-samples, because no
redshift evolution of the \gls{DLA} bias was detected in 
\cite{Perez-Rafols+2018}.

 The green dotted line in the left panel models the measured
distribution of $S$ as an exponential distribution
$P(S)= {\rm e}^{-S/\lambda_{S}}/\lambda_S$, with $\lambda_S=1.2$, convolved
with a Gaussian with error $\epsilon_S=0.6$, which we find fits the
measured distribution adequately. The required error $\epsilon_S$ is
larger than the
statistical errors calculated from the spectra when measuring the
equivalent widths involved in the calculation of $S$, which indicates the
likely presence of unaccounted systematic errors. We will come back to
this question in \cref{sec: error smoothing correction}. The fact that
the true distribution of $S$ can be modelled as an exponential is not
surprising, because the equivalent width distribution of metal lines are
usually reasonably fitted by this form.

\begin{table*}
	\centering	
	\begin{tabular}{cccccc}
		\toprule
		Name & Range & DLA Number & $\overline{S}$ & $b_{\rm DLA}$ & $\chi^{2}\text{ (d.o.f.)}$\\
		\midrule
		$SA$	 & $-2 < S$ & 18,221 & $1.13\pm1.21$ & $2.00\pm0.09$ & 3,001.23 (2,896-1)\\
		$S1$ & $-2 < S < 0.55$ & 6,039 & $0.01\pm0.48$ & $1.67\pm0.16$ & 2,859.52 (2,896-1)\\
		$S2$ & $0.55 \le S < 1.45$ & 5,936 & $0.97\pm0.25$ & $2.03\pm0.16$ & 2,941.40 (2,896-1)\\
		$S3$ & $1.45 \le S$ & 6,246 & $2.44\pm0.91$ & $2.27\pm0.15$ & 2,957.17 (2,896-1)\\
		\midrule
		$C1$ & - & 23,342 & - & $1.97\pm0.08$ & 3,061.04 (2,896-1) \\
		\bottomrule
	\end{tabular}
	\caption{Properties of \gls{DLA} samples $SA$, $S1$, $S2$, and $S3$. The fourth column gives the average value and
dispersion of $S$ in each sample,
computed by weighting the samples with weights equal to $\epsilon_{S}^{-1}$.
	Bias values are given at the reference redshift $z_{\rm ref}=2.3$. Results for sample $C1$ of \protect\cite{Perez-Rafols+2018} are given for reference. The $\chi^2$ parameter of the fit of the cross-correlation in each sample is in the last column.}
	\label{ta DLA: sample properties}
\end{table*}

%
\section{Cross-correlation: Measurement and model}
\label{sec DLA: Cross-correlation}

In this section we summarize our method to compute the cross-correlation
of \glspl{DLA} and the \lya{} forest transmission fluctuation
$\delta_i = 1-F_i/\bar F$ (where $F_i$ is the transmission fraction at
any pixel $i$ of a quasar spectrum and $\bar F$ is its mean value), the
covariance matrix of this cross-correlation, and the model
used to infer the \gls{DLA} bias from a fit to the cross-correlation.

\subsection{Measurement of the cross-correlation}

  We use the same estimator for the cross-correlation of \glspl{DLA} and \lya{}
transmission fluctuation, $\xi$, as in
\cite{Font-Ribera+2012,Perez-Rafols+2018}, where the method is described
in more detail. After dividing the plane of
the parallel and perpendicular components of the separation vector from
a \gls{DLA} to a \lya{} forest pixel, $(r_\parallel,r_\perp)$, into bins that
we designate by $A$, the cross-correlation at the bin $A$ is
\begin{equation}
	\label{eq DLA: xi}
	\xi^{A} = \frac{\sum_{i\in A}w_{i}\delta_{i}}{\sum_{i\in A}w_{i}} ~,
\end{equation}
where the sum is over all \glspl{DLA} and over all pixels $i$ located
within a bin A of the separation {\bf r} from a \gls{DLA} (note that a
given \lya{} forest pixel may appear several times in this sum, as many
as \glspl{DLA} there are within the separation $A$ from the pixel). Here, the
fluctuations $\delta_i$ have been corrected for continuum fitting in an
operation that we call "projection", and
the weights $w_i$ are defined to optimize the accuracy of the measurement
of $\xi^{A}$ \citep[see][and references therein for a detailed description]
{Perez-Rafols+2018}. 

 The covariance of the cross-correlation at two bins $A$ and $B$ is equal to
\begin{equation}
	\label{eq DLA: covariance matrix}
	C^{AB} \equiv \left<\xi^{A}\xi^{B}\right>-\left<\xi^{A}\right> \left<\xi^{B}\right>=\frac{1}{N^{AB}} \sum_{i\in A} \sum_{j\in B} w_iw_j\, \zeta_{ij} ~,
\end{equation}
where $\zeta_{ij}$ is the \lya{} forest autocorrelation at two pixels $i$
and $j$, and each of the two sums is again understood to be over all \lya{}
forest pixels and all the \glspl{DLA} at separations within the bins $A$
or $B$. To compute $C^{AB}$, we include only pairs of \lya{} pixels that
are on the same spectrum, i.e.,
separated only by a parallel component. We neglect the contribution to
the covariance matrix of pixels from different forests.
The normalization factor is
\begin{equation}
	\label{eq DLA: covariance matrix norm factor}
	N^{AB} = \sum_{i\in A} \sum_{j \in B} w_i w_{j} ~.
\end{equation}

\subsection{Model of the cross-correlation}
 \label{subs DLA: Distortion matrix}

We model the cross-correlation starting from the cross-power spectrum, and then Fourier Transform it to obtain the model cross-correlation. We compute the poser spectrum assuming the linear theory of redshift space distortions of
\cite{Kaiser1987} as in \cite{Font-Ribera+2012,Perez-Rafols+2018}:
\begin{equation}
	\label{eq DLA: PS model}
	P_{\rm DLA,Ly\alpha}\left({\bf k}, z\right) = b_{\rm DLA}
 \left(1+\beta_{\rm DLA}\mu_k^2 \right) \, b_{\rm Ly\alpha}
 \left(1+\beta_{\rm Ly\alpha}\mu_k^2 \right) \,
 P_L({\bf k}, z) G({\bf k}) S(k_{\parallel}) ~,
\end{equation}
where $b_{\rm DLA}$ and $b_{\rm Ly\alpha}$ are the bias factors of \glspl{DLA}
and the \lya{} forest, $\beta_{\rm DLA}$ and $\beta_{\rm Ly\alpha}$ their
redshift space distortion parameter, $\mu_k=k_\parallel / k$ is the
cosine of the angle of the Fourier mode vector with respect to the line
of sight, and $P_{L}({\bf k}, z)$ is the linear matter power spectrum.
The functions $G$ and $S$ are added to account for the smoothing that occurs
due to the spectrograph resolution, the binning of the \lya{} forest
spectra, and the binning of the $(r_\parallel, r_\perp)$ plane to measure
the cross-correlation. Details of the assumed form for $G$ and $S$ are
found in section 4 of \cite{Perez-Rafols+2018}. We use comoving bin
sizes $\Delta_{\parallel} = \Delta_{\perp} = 2\hmpc$ to compute the
cross-correlation function, out to maximum separation components
$r_{\parallel} < 80 \hmpc$ and $r_{\perp} < 80 \hmpc$.
In addition, we apply to
the model the same projection operation discussed above that is applied
to the data to correct for continuum fitting effects, as explained in
appendix B.3 of \cite{Perez-Rafols+2018}.

  We use the publicly available code {\it baofit} to fit the model,
which computes the linear power spectrum $P_L$ using CAMB
\citep{Kirkby+2013, Lewis+2011}. The model is evaluated at the mean
values of $r_{\parallel}$ and $r_{\perp}$ of each bin, at the mean
redshift of our sample.

  As in \cite{Perez-Rafols+2018}, we fix
$b_{\rm Ly\alpha}\left(1+\beta_{\rm Ly\alpha}\right)=-0.325$ at the reference redshift $z_{\rm ref}=2.3$,
and $\beta_{\rm Ly\alpha}=1.663$, as found from a detailed
measurement of the \lya{} forest transmission auto-correlation in
\cite{Bautista+2017}. We assume the amplitude of the cross-correlation
evolves proportionally to $\left(1+z\right)^{0.9}$, corresponding to
$b_{\rm Ly\alpha} \propto (1+z)^{2.9}$ as found previously in
\cite{McDonald+2006,Palanque-Delabrouille+2013}, and a constant \gls{DLA}
bias, consistently with the results of \cite{Perez-Rafols+2018}.
We also assume a non-evolving $\beta_{\rm Ly\alpha}$, and we fix
$\beta_{\rm DLA}b_{\rm DLA}=f\left(\Omega\right)=0.9689$.

 In summary, we fix all the parameters in \cref{eq DLA: PS model} except
for the \gls{DLA} bias factor. Our reported errors on $b_{\rm DLA}$
include only measurement errors of the \gls{DLA}-\lya{} cross-correlation, and
not the errors on the \lya{} bias parameters or any other ingredients in
our modelling. Our derived values of $b_{\rm DLA}$ decrease with the
amplitude of $P_L$ and with the value of $b_{\rm Ly\alpha}$ assumed in
our model. However, in this paper we are interested in the variation of
$b_{\rm DLA}$ with the metal strength $S$, which is not
affected by these parameters (except for the small effect caused by the
variation of $\beta_{\rm DLA}$ with $b_{\rm DLA}$, which means that
$b_{\rm DLA}$ is not exactly inversely proportional to the assumed
amplitude of $P_L$ and $b_{\rm Ly\alpha}$).

%
\section{Results and discussion} 
\label{sec DLA: Results}

\begin{figure*}
	\centering
	\includegraphics[width=\textwidth]{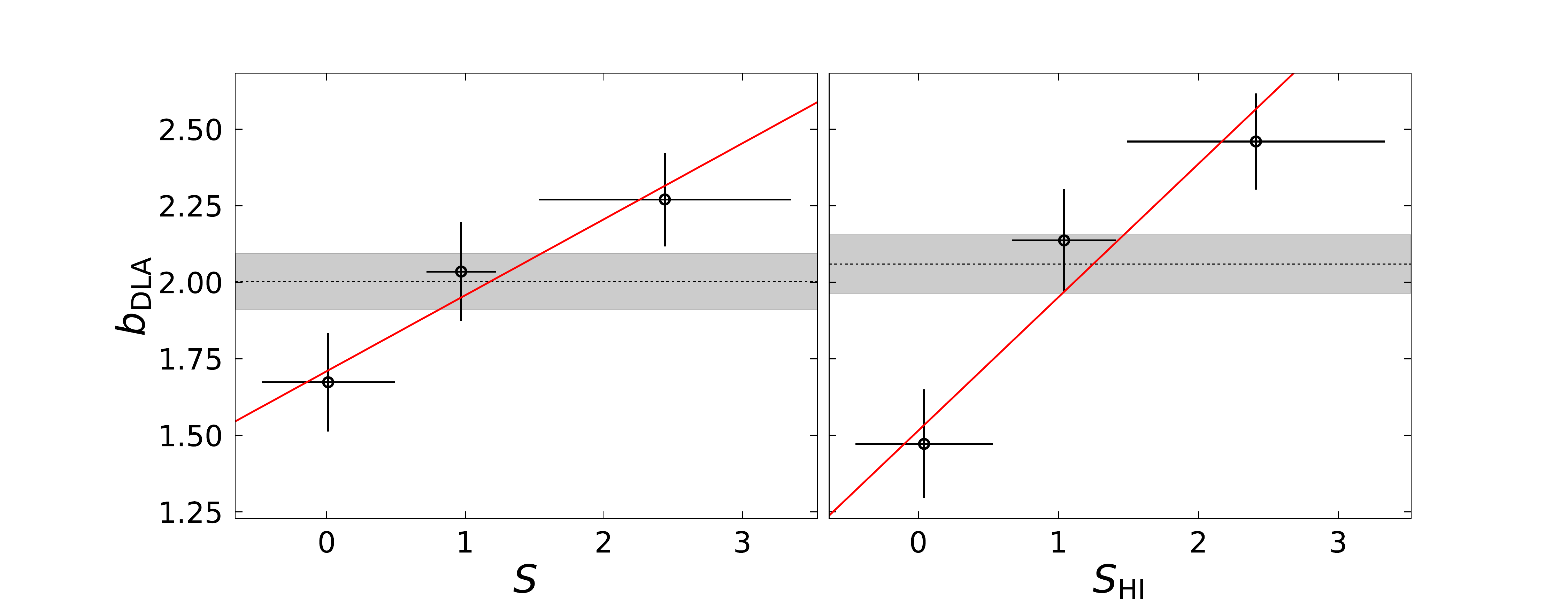}
	\caption{{\it Left}: Bias of the \glspl{DLA} for the three bins
in $S$ of samples $S1$, $S2$, and $S3$ (see \cref{ta DLA: sample properties}),
with errors shown as vertical bars. The value for sample SA is shown by
the black dashed line, with its error indicated by the shaded area. Red
solid line shows a linear fit to the points. {\it Right}: Same as in
left panel for the corrected metal strength $S_{\rm HI}$.}
	\label{fig DLA: biases}
\end{figure*}

 We now present the results of our fits to the measured
cross-correlations for each sample. We exclude from the fit bins at
very small separations, with
$r=\sqrt{r_{\parallel}^{2}+r_{\perp}^{2}} \le 5\hmpc$.
The measured \gls{DLA} bias factors are listed in
\cref{ta DLA: sample properties} and shown in the left panel of
\cref{fig DLA: biases}. We also include the measurement for sample $C1$ of
\cite{Perez-Rafols+2018} in \cref{ta DLA: sample properties} as a
reference value.

  There is a clear increase of the bias factor with $S$. A simple linear
fit, shown by the red solid line, yields
$b_{\rm DLA}(S) = (0.25\pm 0.06 ) S + (1.71\pm 0.09)$, with a slope that
is greater than zero at $4\sigma$. In the rest of this section we
discuss how this relation
changes when we use the metal strength corrected for its dependence on
$N_{\rm HI}$ (\cref{sec DLA: SHI}), the possible impact of impurities on
these results (\cref{sec DLA: impurities}), and the correction we
compute for the bias - metal strength relation that is caused by the
measurement errors in $S$ (\cref{sec: error smoothing correction}).
Finally, we discuss how this result relates to the mass-metallicity
relation of galaxies in \cref{sec:mass-metallicity}.

\subsection{HI-corrected metal strength}\label{sec DLA: SHI}

 The metal strength is expected to increase with the column density of
each metal species and with the velocity dispersion of the absorbing
gas. At the same time, the theory of halo formation in theories with
hierarchical clustering tells us that the bias factor should increase
with the host halo mass. The host halo mass is directly related to
the velocity dispersion, and we also expect a mass-metallicity relation
to exist, as observed in galaxies, implying that more massive halos
should have a higher metal abundance in the gas phase. However, at
fixed metallicity, the column density of metal species should increase
with $N_{\rm HI}$, and should in fact be directly proportional to
$N_{\rm HI}$ if the average ionization conditions in \glspl{DLA} do not
change with $N_{\rm HI}$. Therefore, if these arguments are correct, a
corrected metal strength $S$ that makes it independent of $N_{\rm HI}$
should increase the variation of the bias factor with this parameter.

  We use the corrected metal strength $S_{\rm HI}$ as defined by
\cite{Arinyo-i-Prats+2018}, by empirically determining a linear fit to
the dependence of the average value of $S$ on $N_{\rm HI}$ and
subtracting it from $S$.
We repeat the measurement made above for $S_{\rm HI}$, dividing the
\glspl{DLA} into the same three bins used for $S$. This is justified because the
distributions of $S_{\rm HI}$ and $S$ are very similar (see
\cref{fig DLA: dla hist}). The properties
of the new samples for the corrected metal strength are specified in
\cref{ta DLA: biases corr}. The total number of \glspl{DLA} is lower
when we use $S_{\rm HI}$ because of the cuts $S_{\rm HI} > -2$ and
$\epsilon_{S,{\rm HI}} < 0.5$, which eliminate a larger number of \glspl{DLA}
than for the case when we use $S$. The dependence of the bias on
$S_{\rm HI}$ is shown in the right panel of \cref{fig DLA: biases} and
in \cref{ta DLA: biases corr}.

\begin{table*}
	\centering
	\begin{tabular}{ccccc}
		\toprule
		Name & Number of \glspl{DLA} & $\overline{S}_{\rm HI}$ & $b_{\rm DLA}$ & $\chi^{2} (d.o.f.)$\\
		\midrule
		$SA_{\rm HI}$ & 16,666 & $1.19\pm1.20$ & $2.06\pm0.10$ & 3,041.56 (2,896-1)\\
		$S1_{\rm HI}$ & 4,951 & $-0.01\pm0.51$ & $1.47\pm0.18$ & 2,894.13 (2,896-1)\\
		$S2_{\rm HI}$ & 5,630 & $1.01\pm0.36$ & $2.14\pm0.17$ & 2,947.77 (2,896-1)\\
		$S3_{\rm HI}$ & 5,953 & $2.41\pm0.92$ & $2.46\pm0.16$ & 3,038.79 (2,896-1)\\
		\bottomrule
	\end{tabular}
	\caption{Properties of the \gls{DLA} sub-samples $SA_{\rm HI}$,
$S1_{\rm HI}$, $S2_{\rm HI}$, and $S3_{\rm HI}$, obtained by dividing
the \glspl{DLA} into the same bins in $S_{\rm HI}$ as the ones used for $S$
shown in \cref{ta DLA: sample properties}. The result for the bias factor is
given in the fourth column, and the $\chi^2$ of the fit in the fifth
column.}
	\label{ta DLA: biases corr}
\end{table*}

  The result shows that indeed, the bias factor varies more strongly
with the corrected metal strength. The same linear fit that was obtained
previously for $S$ now gives the result 
$b_{\rm DLA}(S_{\rm HI}) = \left(0.44\pm0.13\right)S_{\rm HI} +
 \left(1.52\pm0.19\right)$, with a substantially steeper slope.
This is consistent with our interpretation that we have detected a
physical dependence of the bias factor with metallicity and velocity
dispersion, and that metallicity is better characterized by the
corrected metal strength $S_{\rm HI}$ than by $S$. Note that, as found
by \cite{Perez-Rafols+2018}, there is no dependence of the bias factor
on $N_{\rm HI}$ within measurement errors, which might have affected
the relation of bias and $S_{\rm HI}$ if it were present.

\subsection{Dependence on the cuts in $\epsilon_{S}$ and
continuum-to-noise ratio}\label{sec DLA: impurities}

  The dependence of the bias factor on $S$ and $S_{\rm HI}$ we have
detected might be contaminated by the lack of purity of the \gls{DLA} catalogue
we use, i.e., the fact that our \glspl{DLA} are only candidates and some of
them may be false \glspl{DLA} arising from spectral noise or from other
absorption systems that are confused with \glspl{DLA}. For example, if a
fraction of \glspl{DLA} in our catalogue were simply arising from noise, their
bias factor and their metal strength would both be zero, creating an
artificial correlation of the bias and the metal strength.

  To test for the possible presence of this contaminating effects,
we check for variations of our results with the cut in the error
$\epsilon_S$ of the metal-strength parameter, and with the
\gls{CNR} in the \lya{} forest region of the
spectrum that the \gls{DLA} is detected in. 

\begin{figure*}
	\centering
	\includegraphics[width=\textwidth]{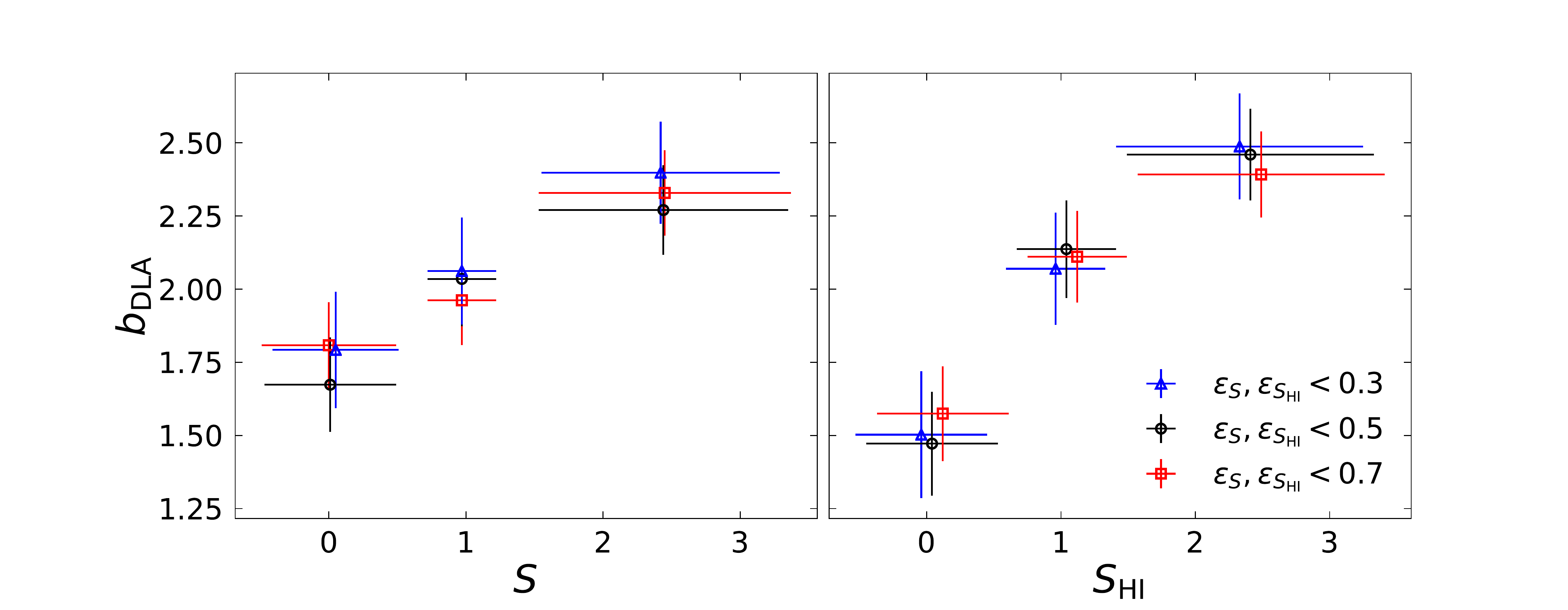}
	\caption{Bias of the \glspl{DLA} against S obtained by fitting
samples $S1$, $S2$, and $S3$ when the cut in $\epsilon$ is changed. Right
panel is for sub-samples in $S_{\rm HI}$. }
	\label{fig DLA: biases sigma S}
\end{figure*}

 We start modifying the cut in $\epsilon_{S}$. As $\epsilon_S$ increases,
the true sub-samples with different values of $S$ are increasingly mixed
in the sub-samples we construct from the measured $S$.
We show in \cref{fig DLA: biases sigma S} the bias as a function of $S$
and $S_{\rm HI}$, for thresholds $\epsilon_S < (0.3, 0.5, 0.7)$. There
is indeed a small reduction of the change in bias with the metal strength
as the maximum allowed $\epsilon$ is raised, that can
be explained by the mixing of sub-samples. However, we do not see a
reduction of $b_{\rm DLA}$ at the smallest $S$ or $S_{\rm HI}$ that
might have been caused by a large fraction of fake \glspl{DLA} among the
sub-sample where metal lines are not significantly detected.

\begin{figure*}
	\centering
	\includegraphics[width=\textwidth]{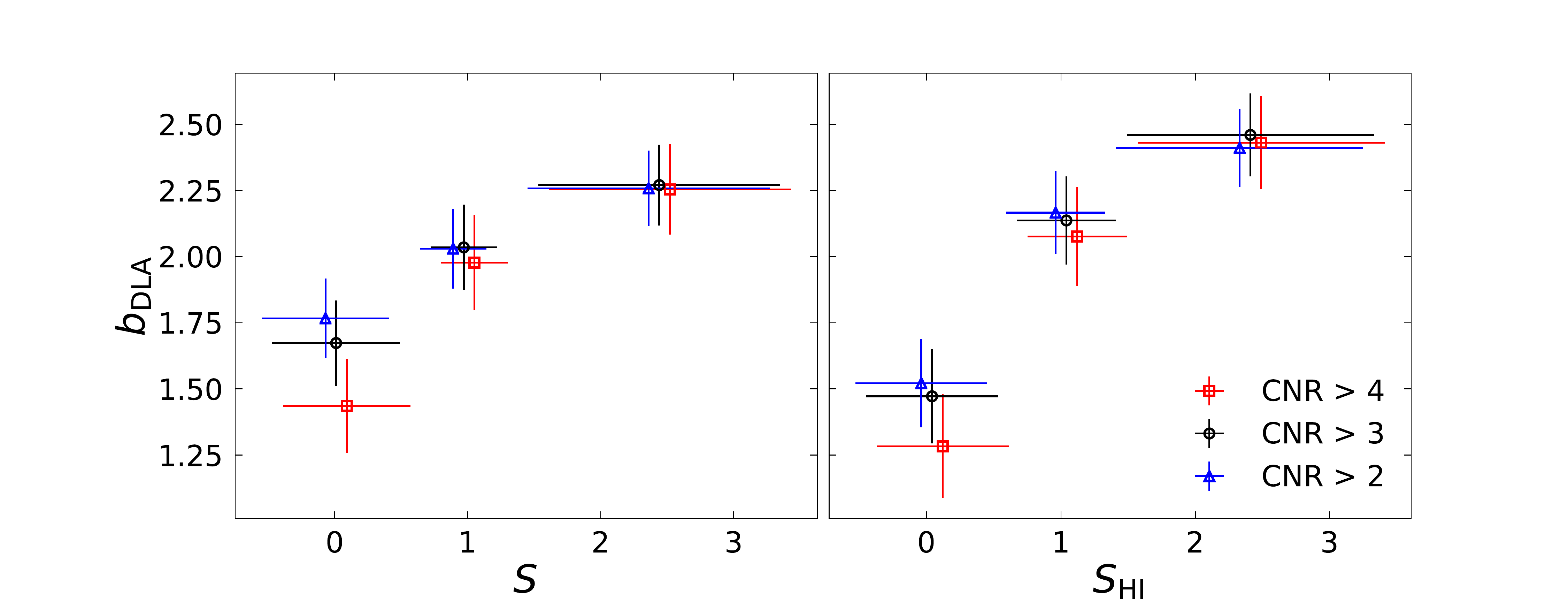}
	\caption{Bias of the \glspl{DLA} against S (left panel) and
$S_{\rm HI}$ (right panel) for different cuts in \gls{CNR}.
Red squares (\gls{CNR} $ > 4$) and blue triangles (\gls{CNR} $ > 2$)
have been horizontally shifted to avoid overlap.}
	\label{fig DLA: biases CNR}
\end{figure*}

  We now move our attention to the dependence of the bias-metal strength
relation on the minimum value required for the continuum-to-noise ratio
of the spectrum in the \lya{} forest region where each \gls{DLA} is detected.
The purity of the catalogue should increase with \gls{CNR}, so any
effect of impurities should decrease rapidly with the threshold we
impose on \gls{CNR}. 
We show this in \cref{fig DLA: biases CNR}, where we see again an
increased variation of \gls{DLA} bias with metal strength as the minimum
required \gls{CNR} is increased.

  Furthermore, in this case we see a larger variation of the bias for
the smallest value of $S$: a higher threshold in \gls{CNR} results in a
lower bias factor for the \glspl{DLA} with the smallest metal strength.
The variation is highly significant: taking into account that the
sub-samples with \gls{CNR} $ > 4$ contain 70\% of the same \glspl{DLA} in the
sub-samples with \gls{CNR} $ > 2$, the expected statistical fluctuation
in the difference of bias factors is $0.4\sigma$, where $\sigma$ stands
for the error in the largest sample, with \gls{CNR} $ > 2$ (indicated by
the blue error bar in the figure). The expected fluctuation comes from the fact that the subsample with \gls{CNR}$ > 2$ can be considered as the combination of two independent samples (one with \gls{CNR}$ > 4$, and another one with $2 < $\gls{CNR} $ \le 4$) that are combined with weight $\sigma_{\rm CNR > 4}^{2}/\sigma_{\rm CNR > 2}^{2}$ to construct the final estimator. This implies that the typical fluctuation is $\delta \sigma_{\rm CNR > 2}$, where $\delta=1-\sigma_{\rm CNR > 4}^{2}/\sigma_{\rm CNR > 2}^{2}$. This variation is opposite to our
expectation if impurities caused only by noise were present in the
catalogue, in which case false \glspl{DLA} would have metal strength and bias
consistent with zero.

  Our interpretation of this result is that the contaminants that are
included in the catalogue as the \gls{CNR} threshold is lowered are mostly
regions of broad \lya{} absorption in the forest, which correspond to
absorbers of lower column density than \glspl{DLA} but with sufficiently broad
velocity dispersions to create a broad absorption line that
is consistent with a \gls{DLA} in the most noisy spectra. These absorbers
generally have weaker associated metal lines because of their lower
column densities and higher ionization level
\citep[most of the lines used to
measure the metal strength are of low-ionization species, see][]{Arinyo-i-Prats+2018}, and they are more highly biased than
the \glspl{DLA} with the weakest metal lines because they are associated with
collapsed regions, or regions in the process of collapsing, with high
velocity dispersion. This can explain why the presence of contaminants
increases the bias factor, and actually decreases the variation we
measure of the bias factor with metal strength. These type of absorbers
and their average metal lines are likely similar to the systems studied
by \cite{Pieri+2014}; a detailed study of the bias factor of these strong \lya{}
absorbers is separately being done \citep{Blomqvist+Inprep}.

\subsection{Correction for the effect of metal strength measurement errors}
\label{sec: error smoothing correction}

  The errors in the measurement of $S_{\rm HI}$ mix the samples
$S1_{\rm HI}$, $S2_{\rm HI}$ and $S3_{\rm HI}$, as each \gls{DLA} is
classified to one of these samples based on the measured $S_{\rm HI}$.
This mixing among the samples should flatten the dependence of the
\gls{DLA} bias on $S_{\rm HI}$. Hints of this effect are seen in
\cref{fig DLA: biases sigma S}. In this subsection we compute a
correction for this effect based on the distribution of $S$.

  The intrinsic distribution of $S$ can be modelled as an exponential,
\begin{equation}
	\label{eq: P(S) intr}
	P_{\rm in}(S) = {\rm e}^{-S/\lambda_{S}}/\lambda_{S} ~,
\end{equation}
where $\lambda_S$ is a characteristic value for $S$. To take into
account the effect of errors, we convolve this intrinsic distribution
with a Gaussian to obtain the observed distribution,
\begin{equation}
	\label{eq: P(S)}
	P(S) = \int_{-S}^{\infty} d\delta_{S}\,
 P_{\rm in}(S+\delta_{S})\frac{{\rm e}^{-\delta_S^{2}/2\epsilon_{S}^{2}}}{\sqrt{2\pi\epsilon_{S}^{2}}} ~.
\end{equation}
The distribution with parameters $\lambda_{S}=1.2$ and
$\epsilon_{S}=0.6$ reasonably reproduces the observed distribution as
shown in the left panel of \cref{fig DLA: dla hist}. This
good match is not surprising, since we know that the equivalent width
distribution of metal lines is reasonably fitted by an exponential
function, and $S$ was defined to make its average be close to unity.
The typical error in this Gaussian is somewhat larger than the actual
errors reported in the catalogue
\citep[see figure 5 of][]{Arinyo-i-Prats+2018}.
This indicates that the true errors may be larger than the estimate
calculated in \cite{Arinyo-i-Prats+2018} from the statistical errors of
the flux in spectral pixels, probably because of continuum fitting and
other systematic errors. The correction presented here takes this
increased errors into account by calibrating them from the observed
distribution of $S$.

We now use this distribution to compute a correction for the measured
\gls{DLA} bias, given an intrinsic or true relation $b_{\rm DLA,in}(S)$:
\begin{equation}
	\label{eq: bias error correction}
	b_{\rm DLA}(S) = \frac{1}{\mathcal{N}}
\int_{-S}^{\infty} d\delta_{S}\, b_{\rm DLA,in}
\left(S+\delta_{S}\right)\, {\rm e}^{-\left(S+\delta_{S}\right)/\lambda_{S}}\,
\frac{{\rm e}^{-\delta_S^{2}/2\epsilon_{S}^{2}}}{\sqrt{2\pi\epsilon_{S}^{2}}}
 ~,
\end{equation}
where
\begin{equation}
	\mathcal{N} = \int_{-S}^{\infty}d\delta_{S}\,
 {\rm e}^{-\left(S+\delta_{S}\right)/\lambda_{S}}
\frac{{\rm e}^{-\delta_S^{2}/2\epsilon_{S}^{2}}}{\sqrt{2\pi\epsilon_{S}^{2}}}
 ~.
\end{equation}

\begin{figure*}
	\centering
	\includegraphics[width=0.45\textwidth]{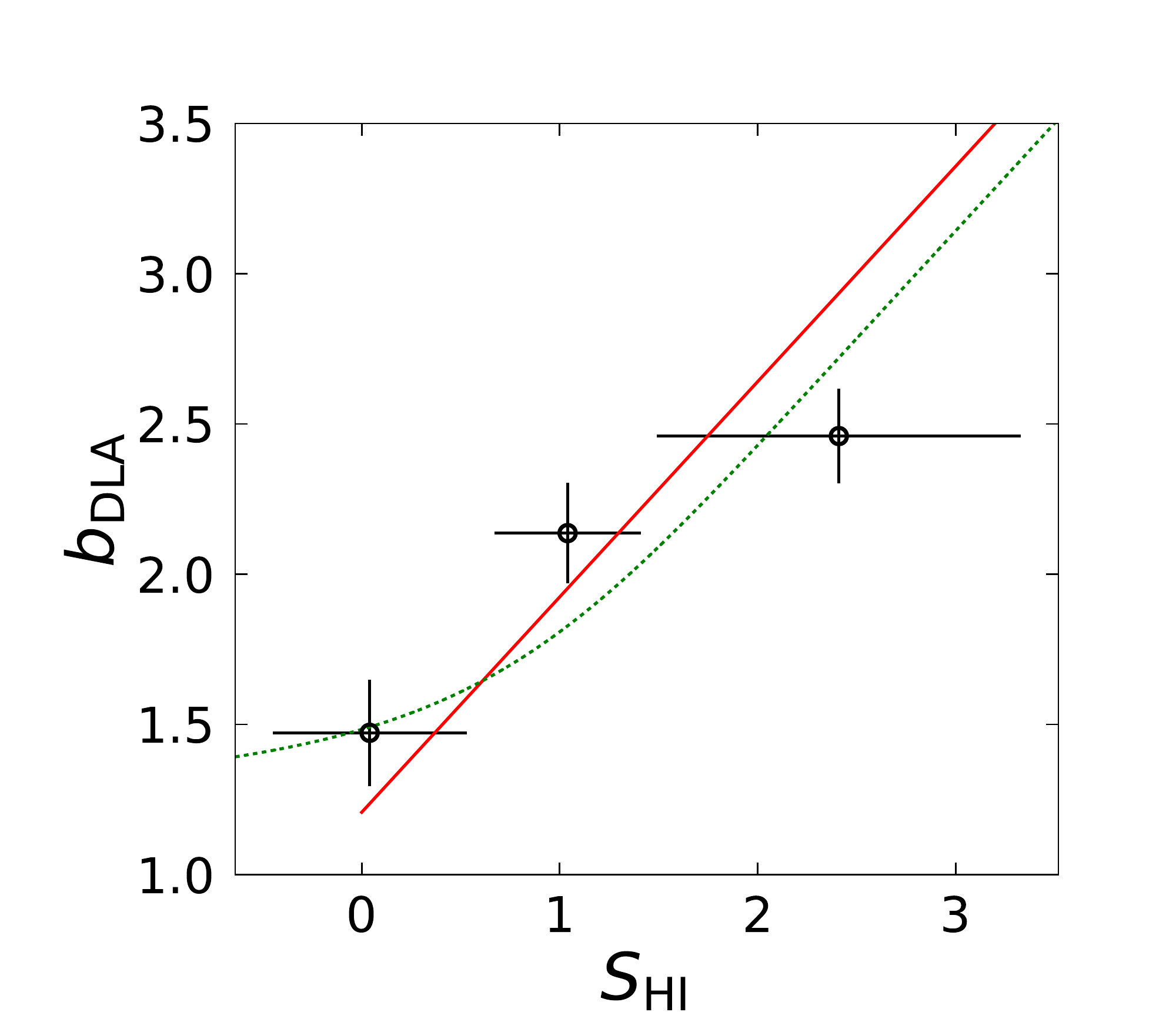}
	\caption{Bias of \glspl{DLA} versus HI-corrected metal strength,
$S_{\rm HI}$. Green dotted line shows a fit to the points assuming an
intrinsic linear relation for the bias, shown as the red solid line,
and applying \cref{eq: bias error correction}.}
	\label{fig DLA: biases correction}
\end{figure*}

 We apply this correction to the measurement of the bias using
$S_{\rm HI}$. We assume that the intrinsic bias-metal strength relation
is linear, and we fit the predicted observed relation to the three
values of our chosen bins. The result is shown in
\cref{fig DLA: biases correction}. The best fit solution is
$b_{\rm DLA\,in}(S) = (0.72\pm 0.30) S_{\rm HI} + (1.21\pm 0.24)$,
and the corresponding observed relation is the green dotted line.
The predicted observed relation has a positive second derivative,
whereas the observed points show a negative second derivative, although
this is not highly significant. In any case, there is no reason why the
intrinsic relation should be linear, and the true relation can only be
predicted from detailed cosmological simulations of the halos giving
rise to \glspl{DLA}, but more data and an increased number of bins would be
necessary to measure the shape of this relation more accurately.
Nevertheless, the simple correction we have applied highlights two
important results. First, the true relation of the \gls{DLA} bias to the
metal strength is substantially steeper than what we have directly
measured because of the spreading effect of measurement errors. Second,
the intrinsic estimated value of the \gls{DLA} bias at zero metal
strength is $1.21\pm0.24$, and is consistent with unity, which is the
value expected for low-mass halos hosting dwarf galaxies 
\citep[$M\sim 10^9$ to $10^{10}\, M_\odot$; see figure 8 of][]{Perez-Rafols+2018}.
This is consistent with the observed mass-metallicity relation, where
more metal poor galaxies have lower stellar mass \citep{Maiolino+2008},
and suggests that \glspl{DLA} are hosted by halos with a very broad range of
masses, with metal-line strength strongly correlating with host halo
mass. 

\subsection{Implications on the mass-metallicity relation}\label{sec:mass-metallicity}

 The relation of the \gls{DLA} bias with the metal strength can arise
from an increase of the \glspl{DLA} metallicity with host halo mass, or
also an increase of the velocity dispersion. The theory of Cold Dark
Matter for structure formation makes a clear prediction of how the
velocity dispersion must increase with halo mass, although uncertainties
can be introduced in the velocity dispersion of the gas depending on its
density profile. The relation of metallicity to stellar mass has been
empirically found in galaxies, and can be measured at present even at
the relatively high redshifts of our \gls{DLA} sample
\citep[e.g.,][]{Ledoux+2006, Maiolino+2008, Moller+2013}. We expect that future studies will be
able to discern the contribution of the velocity dispersion and
metallicity correlations with halo mass by comparing the mean metal
strength of several lines in our different samples
\citep[see][]{Mas-Ribas+2017}, or from individual measurements at high
resolution and signal-to-noise of an adequate sample of \glspl{DLA} to
clarify how velocity dispersion and metallicity affect the metal
strength parameter we have used \citep[see e.g.,][]{Prochaska+2008}.
This promises to allow precise measurements of a mass-metallicity
relation for the gas phase, in terms of halo mass instead of stellar
mass.

%
\section{Summary and conclusions}
\label{sec DLA: Conclusions}

 In this paper, we make use of the metal strengths obtained for a large
number of \glspl{DLA} in \cite{Arinyo-i-Prats+2018} to measure the bias
of \glspl{DLA} as a function of metal strength. We divide the total
\gls{DLA} sample into three subsamples of different metal strength, and
we measure the cross-correlation of each subsample with the \lya{}
forest. We fit a linear theory model to the cross-correlations to derive
the bias factor of \glspl{DLA} as a function of $S$, and as a function
of $S_{\rm HI}$, the metal strength corrected for the dependence on
$N_{\rm HI}$ \citep{Arinyo-i-Prats+2018}
Our main results are summarized as follows:

 \begin{itemize} 
	\item We find a clear dependence of the \gls{DLA} bias on $S$.
A linear fit yields
$b_{\rm DLA}(S) = \left(0.25\pm0.06\right)S + \left(1.71\pm0.09\right)$. For the first time, we find a dependence of the bias factor of \glspl{DLA} with their metal content.
        \item The dependence on $S_{\rm HI}$ is even stronger,
$b_{\rm DLA}(S_{\rm HI}) = (0.44\pm0.13)S_{\rm HI} + (1.52\pm0.19)$, which
confirms that the effect we detect is real and is related to metallicity
rather than metal column density. These detections are statistically
significant at more than $3\sigma$.
	\item We note that the presence of contaminants, which are likely
strong \lya{} absorbers that are confused with \glspl{DLA} at low
signal-to-noise, can
significantly increase the value of the bias for the lowest bin in
$S$ or $S_{\rm HI}$, further increasing the slope of the true linear
relation.
	\item The large errors in the measurement of the metal strength
that are inevitable in our large sample of \glspl{DLA} observed at low
signal-to-noise causes a spreading over $S$ that dilutes the dependence
of bias on metal strength. Calculating a simple correction for this
spreading effect, we fit a linear relation and obtain
$b_{\rm DLA\,in}(S_{\rm HI}) = (0.73\pm0.31)S_{\rm HI} + (1.21\pm0.24)$.
This suggests that the lowest metallicity \glspl{DLA} have a bias close
to unity, which is characteristic of the lowest mass halos hosting dwarf
galaxies, and that the average DLA bias of $b_{DLA}=1.98 \pm 0.08$
found in \cite{Perez-Rafols+2018} is the result of averaging the low
bias of the low-metallicity \glspl{DLA} in dwarf galaxies with \glspl{DLA} in very
massive halos, with high metallicity and velocity dispersion, and bias
factor as high as $2.5$ to 3. Physically motivated models for the true
bias-metal strength relation from cosmological simulations will be
highly valuable in the future.
	\item The bias-metal strength relation we have investigated for
\glspl{DLA} is most likely related to the observed stellar mass-metallicity
relation in galaxies \citep{Maiolino+2008}. We will investigate the
relation between metal strength and metallicity in a future paper.
More complete studies promise to measure a relation of the metallicity
in the gas phase with halo mass, which can help establish the physical
origin of this relation both in the star and gas contents of galaxies.

 \end{itemize}

\vspace{6mm}

\section*{Acknowledgments}
Funding for SDSS-III has been provided by the Alfred P. Sloan Foundation, the Participating Institutions, the National Science Foundation, and the U.S. Department of Energy Office of Science. The SDSS-III web site is http://www.sdss3.org/. 

SDSS-III is managed by the Astrophysical Research Consortium for the Participating Institutions of the SDSS-III Collaboration including the University of Arizona, the Brazilian Participation Group, Brookhaven National Laboratory, University of Cambridge, Carnegie Mellon University, University of Florida, the French Participation Group, the German Participation Group, Harvard University, the Instituto de Astrofisica de Canarias, the Michigan State/Notre Dame/JINA Participation Group, Johns Hopkins University, Lawrence Berkeley National Laboratory, Max Planck Institute for Astrophysics, Max Planck Institute for Extraterrestrial Physics, New Mexico State University, New York University, Ohio State University, Pennsylvania State University, University of Portsmouth, Princeton University, the Spanish Participation Group, University of Tokyo, University of Utah, Vanderbilt University, University of Virginia, University of Washington, and Yale University.

IPR was supported by the A*MIDEX project (ANR-11-IDEX-0001-02) funded by the ``Investissements d'Avenir'' French Government program, managed by the French National Research Agency (ANR), and by ANR under contract ANR-14-ACHN-0021.

AIP, and JME were supported by the Spanish MINECO under
projects AYA2012-33938 and AYA2015-71091-P and MDM-2014-0369 of ICCUB (Unidad de Excelencia 'María de
Maeztu').

AFR acknowledges support by an STFC Ernest Rutherford Fellowship, grant reference ST/N003853/1

\bibliographystyle{apj}
\bibliography{iprafols}

\appendix

\end{document}